\documentclass[twocolumn,showpacs,amsmath,amssymb,prb]{revtex4}

\usepackage{graphicx}
\usepackage{dcolumn}
\usepackage{bm}
\usepackage{psfrag}

\usepackage{color}
\definecolor{red}{rgb}{0.8,0.0,0.0}

\newcommand{\up}{\uparrow}
\newcommand{\dn}{\downarrow}

\newcommand{\bra}[1]{\langle #1|}
\newcommand{\ket}[1]{|#1\rangle}

\newcommand{\aver}[1]{\langle #1 \rangle}
\DeclareMathOperator{\Tr}{Tr}

\begin{document}

\title{Faraday-rotation fluctuation spectroscopy with static and oscillating magnetic fields}

\author{Matthias Braun}
\author{J\"urgen K\"onig}
\affiliation{Institut f\"ur Theoretische Physik III, Ruhr-Universit\"at Bochum, 44780 Bochum, Germany}

\date{\today}

\begin{abstract}
By Faraday-rotation fluctuation spectroscopy
one measures the spin noise via Faraday-induced
fluctuations of the polarization plane of a
laser transmitting the sample. In the fist part
of this paper, we present a theoretical model of
recent experiments on alkali gas vapors and
semiconductors, done in the presence of a {\em static}
magnetic field. In a static field, the spin noise
shows a resonance line, revealing the Larmor
frequency and the spin coherence time $T_2$ of
the electrons.  Second, we discuss the possibility
to use an {\em oscillating} magnetic field in the
Faraday setup. With an oscillating field applied,
one can observe multi-photon absorption processes
in the spin noise. Furthermore an oscillating field
could also help to avoid line broadening due to
structural or chemical inhomogeneities in the sample,
and thereby increase the precision of the
spin-coherence time measurement.
\end{abstract}

\maketitle

\section{\label{Introduction}Introduction}

The anticipation of an application in quantum information
processing\cite{Loss1998,Zutic2004} motivates the intensive
research on spin dynamics and spin coherence, especially in
semiconductors.\cite{Khaetskii2002,Fujisawa2002,Kouwenhoven2004}
Beside orthodox magnetic resonance experiments,\cite{Hahn1950, Blume1958, Kalin2003}
mainly two optical measurement schemes for the spin dynamics are in use:
studies of the Hanle effect,\cite{Epstein2001,Katzer2002}
and time-resolved Faraday (or Kerr) rotation.\cite{Gupta1999,Stern2005}
The former relies on the decrease of photoluminescence polarization
due to an external magnetic field, the latter
on the dependence of the phase velocity of polarized
light on the spin orientation in the sample.
While the Hanle setup measures the spins of
excited electrons, the Faraday setup does not require
this excitation.
That optical measurements are possible at all is a
consequence of spin-orbit coupling.

\begin{figure}[t]
\includegraphics[angle=0,width=\columnwidth]{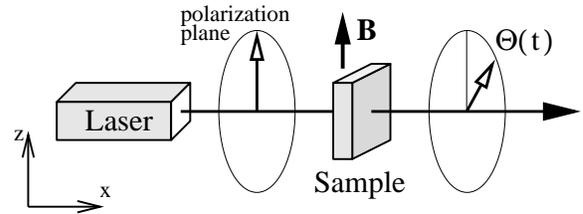}
\caption{\label{fig:setup}
A linear polarized laser is send through the
sample. Due to spin-orbit coupling and the Faraday effect,
the laser polarization plane rotates depending on the
electron spins inside the sample. While the time averaged
Faraday-rotation angle is zero, since there is no net
magnetization collinear to the laser propagation direction, the
Faraday-rotation fluctuations are finite.}
\end{figure}

The time-resolved Faraday rotation has proven
to be a very precise experimental tool, capable
to address the spin dynamics down to the ps time
scale. However, the necessary time resolution for
such experiments requires high experimental efforts,
i.e. a Streak-camera system.
Already in 1981, Aleksandrov and Zapassky
experimentally demonstrated,\cite{Aleksandrov1981}
that instead of measuring the Faraday-rotation in
an alkali gas, one can also measure Faraday-rotation
fluctuations to observe the spin dynamics. Thereby
latter experimental setup which does
not require a time-resolution at all.
They applied a static external magnetic field and transmit linear
polarized laser light through the sample, perpendicular to the
field, see Fig.\ref{fig:setup}. Due to the Faraday
effect, the spin noise in the sample
is mapped on the fluctuations of the laser polarization plane, and
latter can easily be measured. The
precession of the spins in the sample gives rise to a Lorentzian
line in the power spectrum of the Faraday-rotation fluctuations.
The noise frequency of this line is therefore a direct measure of
the Larmor frequency $\omega_0$, and the line width indicates the
electron-spin coherence time $T_2$.

Recently Crooker {\em et al.}\cite{Crooker2004} presented an
increase of the precision of this optical magnetic resonance
experiment\cite{Mitsui2000} capable to
resolve even the different isotope lines in a Rubidium gas.
On the other side Oestreich {\em et al.}\cite{Oestreich2005}
demonstrated that Faraday-rotation fluctuation spectroscopy can
also be used in a solid-state environment: they addressed the
electron spin precession in bulk GaAs.

In the following Sec.~\ref{sec:theory}, we present
an elementary theoretical description of these
existing Faraday-rotation fluctuation experiments\cite{Crooker2004,Oestreich2005,Aleksandrov1981,Mitsui2000}
(that involves a static magnetic field) within a density-matrix formulation.
Then, we use this language to propose a different measurement scheme, namely
to apply an oscillating magnetic field.
In this case, the resonance will not appear at the
(eventually locally varying) {\it Larmor frequency},
but at multiples of the {\it oscillation frequency} of the magnetic field. Therefore some sources of inhomogeneous broadening could be avoided.

\section{\label{sec:theory}Theoretical description}

\begin{figure}[ht]
\includegraphics[width=0.8\columnwidth]{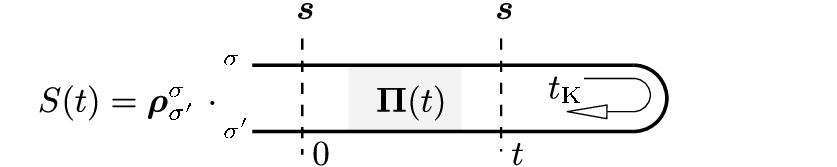}
\caption{\label{fig:diagram}
Diagrammatic representation of $S(t)$. The operators $\hat{s}_{\rm x}$ are
placed on the upper or lower branch of the time contour at time $0$ and $t$,
respectively.}
\end{figure}

Since Faraday rotation measures the spin component collinear with the
laser propagation direction, depending on the optical selection rules
of the sample, the fluctuation of the
Faraday rotation is a direct measure of the
transverse spin-spin correlation function
\begin{eqnarray}\label{correlator}
S(t) &=& \aver{\hat{s}_{\rm x}(t)\hat{s}_{\rm x}(0) +
\hat{s}_{\rm x}(0)\hat{s}_{\rm x}(t)}\,.
\end{eqnarray}
The average of operators can be expressed
by the trace
$\aver{\hat{s}_{\rm x}(t)\hat{s}_{\rm x}(0)}=
\Tr[\hat{s}_{\rm x}(t)\hat{s}_{\rm x}(0)\rho]\,,
$ where  $\rho$ is the SU(2) density
matrix describing one localized
spin in the sample, and
$\hat{s}$ is the $2\times2$ Pauli spin operator.
The time evolution of the operator
$\sigma_x$ in the Heisenberg picture reads
$\hat{s}_{\rm x}(t)=(\hbar/2)
\exp[ i/\hbar \int^t_0 \hat{H}(\tau)d\tau]
\sigma_x
\exp[-i/\hbar \int^t_0 \hat{H}(\tau)d\tau]$, if
Hamiltonians at different times commute, which
will be the case for a magnetic field with static or
oscillating field strength.
It is convenient to represent the correlator $S(t)$ as a diagram with the
operators $\hat{s}_{\rm x}(0)$ and $\hat{s}_{\rm x}(t)$ placed on a Keldysh
time contour $t_{\rm K}$, see Fig.~\ref{fig:diagram}.
The spin states $\uparrow,\downarrow$
propagate along the upper time contour from
the time $0$ to $t$, and on the lower line from
$t$ to $0$. This takes into account the Heisenberg
time evolution of $\hat{s}_{\rm x}(t)$.

Instead of reading the time contour along $t_{\rm K}$
for the individual spin states of the Hilbert space, it is more
intuitive to read the diagram from left to right,
and interpret the double line of upper and lower Keldysh propagator as the
time evolution of the whole density matrix ${\rho}^{\sigma}_{\sigma^\prime}$.
We start with an initial density matrix at time
$0$, and measure the spin state by the operator
${\bm s}$.  Then, the density matrix propagates
from time $0$ to time $t$, and the spin is measured again.
From this perspective, Eq.~(\ref{correlator}) can be rewritten as
\begin{eqnarray}\label{trace}
S(t)&=& \frac{1}{2}\Tr[{\bm s}\,{\bm \Pi}(t)\, {\bm s}\, { \rho}]\,.
\end{eqnarray}
The (Liouville) operator $\bm s$ accounts for placing $\hat{s}_{\rm x}$
at the upper or lower Keldysh contour, flipping the spin from $\sigma$
to $\bar \sigma$.
By summing over both possibilities, the time symmetrization of
Eq.~(\ref{correlator}) is taken into account\cite{note1}.
The (Liouville) operator $\bm s$ is a fourth-order tensor with
$s_{\sigma^\prime\sigma^\prime}^{\sigma\hphantom{^\prime}\bar{\sigma}}
={s}^{\sigma^\prime\sigma^\prime}_{\sigma\hphantom{^\prime}\bar{\sigma}}
=\hbar/2$ and zero otherwise.

Between the two spin measurements at time $0$ and time $t$,
the propagation of the density matrix is described by
$\Pi_{\sigma^\prime\sigma^\prime}^{\sigma\hphantom{^\prime}\sigma}(t)=
\exp \left\{ -i/\hbar\int^t_0 [\bra{\sigma}\hat{H}(\tau)\ket{\sigma}
-\bra{\sigma^\prime}\hat{H}(\tau)\ket{\sigma^\prime} ] d\tau \right\}$.
Since we assume the magnetic field to be along the
spin quantization axis, the states $\sigma, \sigma'$ are eigenstates of the
Hamilton operator, and no elements of $\Pi$ with different spin indices
on the upper (or lower) propagator appear.

The formulation Eq.~(\ref{trace}) offers the possibility to phenomenologically
include transverse spin relaxation in the calculation.
If the initial density matrix was in a diagonal state, the first spin operator
$\hat{s}_{\rm x}$ brings it in a non-diagonal state
$\rho^\sigma_{\bar{\sigma}}$. During the time evolution from $0$ to $t$
non-diagonal density matrix elements decay exponentially with the
time scale set by the spin-coherence time $T_2$. This is accounted for by multiplying
$\Pi^{\sigma\sigma}_{\bar{\sigma}\bar{\sigma}}(t)$ with $\exp(-t/T_2)$.
Such a exponential decay can for example be justified within
a white noise model.\cite{Berns2006}

\section{\label{static}Static magnetic field}

\begin{figure}[h!]
\includegraphics[angle=-90,width=0.9\columnwidth]{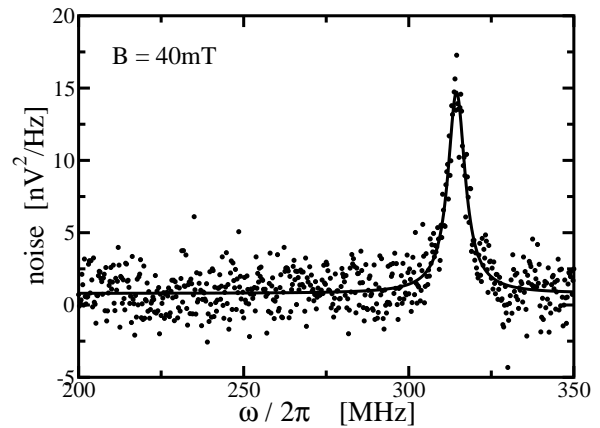}
\caption{\label{fig:experiment}
Measured Faraday-rotation
noise of a spin ensemble by Oestreich {\em et al.},\cite{Oestreich2005}
together with a fit of a Lorentzian function as expected for a single spin,
see Eq.~(\ref{result}).}
\end{figure}

To apply the formulation introduced above to the experiments presented in Ref.~
\onlinecite{Crooker2004,Oestreich2005,Aleksandrov1981,Mitsui2000}, we set
$ H(t)=- \omega_0\, \hat{s}_{\rm z}$ with $\omega_0=g\mu_{\rm B} B/\hbar$.
Consequently, the propagator for a spin is then given by
%
\begin{eqnarray}\label{propagator}
\Pi^{\sigma\sigma}_{\sigma\sigma}=1\,,\,\,\,\,\,
\Pi^{\up\up}_{\dn\dn}=e^{i \,\omega_0 t - t/T_2}\,,\,\,\,\,\,
\Pi_{\up\up}^{\dn\dn}=e^{-i \,\omega_0 t - t/T_2} \,,
\end{eqnarray}
and zero otherwise.
The resulting power spectrum $S(\omega)= \int_{0}^\infty dt
[\exp (-i\omega t) + \exp (i\omega t) ] S(t)$ of the time-dependent
correlator $S(t)$ equals then
\begin{equation}\label{result}
S(\omega) = \frac{\hbar^2}{2}\left(\frac{T_2}{1+{T_2}^2(\omega-\omega_0)^2}+\frac{T_2}{1+{T_2}^2(\omega+\omega_0)^2}\right)\, .
\end{equation}
It shows a Lorentzian resonance centered at the Larmor frequency
$\omega_0$, and the resonance width is given by $T_2$, comparable to a
continuous-wave ESR experiment.
We performed a fit to the experimental data of Oestreich
{\em et al.}\cite{Oestreich2005} in Fig.~\ref{fig:experiment}.

\begin{figure}[h!]
\includegraphics[width=1\columnwidth]{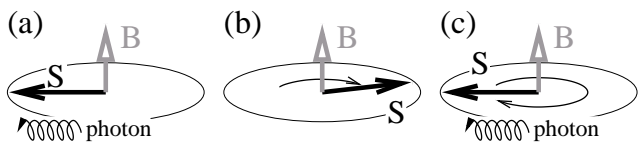}
\caption{\label{fig:diagram1}
Sketch of the electron spin dynamics in the Faraday setup
with a static magnetic field. (a) At time $t=0$,
the spin is measured. (b) Between the two measurement, the
spin precesses in a static magnetic field $\bm B$.
(c) If the second measurement takes place after a full
revolution, it reproduces the outcome of the first measurement.}
\end{figure}

The origin of the resonance can be understood by tracking the time
evolution of the spin between two measurements.
Here for simplicity, we consider only the spin
component along the laser propagation
direction, i.e., along the $x$-axis, since only this component
determines the Faraday rotation of the laser polarization plane.
Let us assume that at time $t=0$ the
spin is aligned parallel to the direction of measurement,
as shown in Fig.~\ref{fig:diagram1}. The outcome of the first
measurement therefore equals $s_{\rm x}=+\hbar/2$.
This spin then precesses in the static external magnetic field.
If the time between first and second measurement is an
integer times the time for a full revolution of the spin, the outcome
of the second measurement will for certain be
$s_{\rm x}=+\hbar/2$, i.e., the results of the two measurements coincide.
The probability to measure the same spin state again is reduced 
by spin relaxation / decoherence.

For half-integer multiples, the measurement results will have opposite signs.
The spin-spin correlation function is, therefore, an oscillating function in
time, which is exponentially damped due to decoherence. The corresponding
power spectrum is a Lorentzian centered at the
Larmor frequency $\omega_0$ with a width of $T_2$.
This argument also holds, if the initial spin state is a coherent
superposition of spin up and down, i.e., the initial spin
is unimportant. In contrast to an ESR experiment,
no net magnetization of the semiconductor sample is needed,
which is reflected in the fact, that Eq.~(\ref{result}) is
independent of the initial density matrix.

\section{\label{oscillating}Oscillating magnetic field}

\begin{figure}[h]
\includegraphics[width=0.97\columnwidth]{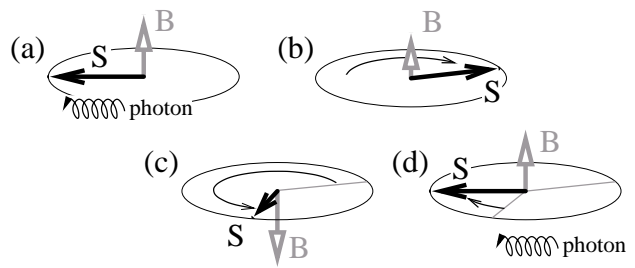}
\caption{\label{fig:diagram2}
Sketch of the spin dynamics in the Faraday
setup with an oscillating field:
(a) The initial measurement of the spin state.
(b) In the external field the spin precesses in one direction.
(c) When the field changes its sign, the spin precesses in the opposite
direction.
(d) After a full field-oscillation time, the spin is again in its initial
state.}
\end{figure}

We now turn to the case of an oscillating
magnetic field along the $z$-direction.
The difference to the case of a static magnetic
field is that the spin precession between the
two measurements can change its direction as a
function of time. This is quite similar to the case in
spin-echo resonance experiments.\cite{Hahn1950,Blume1958}
After the first measurement the spin
precesses in one direction but then stops and
precesses back. After a full oscillation period
of the external field, the spin will be just
back at its starting point, see Fig.~\ref{fig:diagram2}.
If the second measurement takes place at this
time, the outcome will most probable be equal
to the first measurement, i.e., it will be correlated.
The probability of equal spin measurements is
decreased by decoherence.

\begin{figure}[h!]
\includegraphics[angle=-90,width=0.95\columnwidth]{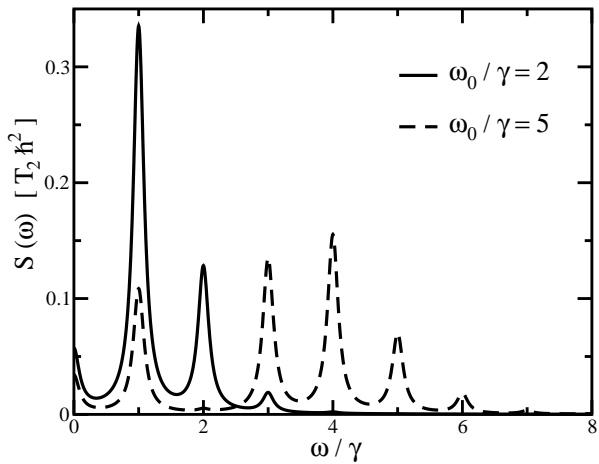}
\caption{\label{fig:plotomega}
Spectrum of the spin-spin correlation function
for different ratios of magnetic field
strength to oscillation frequency.
For a higher ratio $\omega_0/\gamma$,
more resonance lines appear at higher
frequencies, while the signal strength
decreases.
Here, we chose $T_2^{-1}=0.1\gamma$.}
\end{figure}

Technically, we can describe the oscillating magnetic
field by the time-dependent Hamiltonian
\begin{eqnarray}
H(t)=-\omega_0 \cos(\gamma t +\phi)\cdot \hat{s}_z
\end{eqnarray}
with the field oscillation frequency $\gamma$. With this
Hamiltonian, we can derive the
propagator $\bm \Pi(t)$, including a phenomenological
relaxation term. Since the experiment should be a
continuous-wave experiment, there shall be no
correlation between the phase of the magnetic field
oscillation, and the absolute time of measurement of the
spins. Therefore, we average over the
phase $\phi$, and get $\Pi^{\up\up}_{\dn\dn}(t)= \int_0^{2\pi} (d\phi/2\pi)
\exp[-t/T_2 +i \omega_0 / \gamma \,(\sin(\gamma t +\phi)-\sin \phi )]$.
Making use of the Jacobi-Anger expansion,\cite{Abramowitz} and the integral
representation of the Bessel function, we obtain the power spectrum of the
spin-spin correlation function as
\begin{eqnarray}\label{result2}
S(\omega)\,=\hbar^2 \sum_{n=-\infty}^{+\infty}
\left[  J_n\left( \frac{\omega_0}{\gamma} \right) \right]^2
\,\frac{T_2}{1+{T_2}^2(\omega+n \gamma)^2}\,.\,\,\,
\end{eqnarray}
The power spectrum consists of a series
of Lorentzian resonances of the width
$T_2^{-1}$ as plotted in Fig.~\ref{fig:plotomega}.
The different Lorenzian lines in the noise spectrum
correspond to multi-photon absorption processes. Several
photons of the exciting magnetic field are converted in 
a single noise quantum with correspondingly increased frequency.
The appearance of the multi-resonances is extensively
discussed in the context of ESR and NMR, see
Ref.~\onlinecite{Kalin2003} and citations therein.

The signal strength of the Faraday spectrum is proportional to the
square of Bessel functions $J_n$ of the first kind.
If the argument of these Bessel functions is
of the order of $1$, one can expect
approximatively half of the signal magnitude compared to
the case of a static magnetic field.\cite{Oestreich2005} Therefore, every experimental setup capable to resolve Faraday-rotation
fluctuation in a static field should also do
so in an oscillating field.

The property of the correlation spectrum, that the
resonances appear at multiples of the oscillation 
frequency $\gamma$ of the external magnetic field
rather than the Larmor frequency $\omega_0$ offers the
possibility to utilize such a measurement for metrology.
Since the field oscillation frequency $\gamma$ does not vary
locally over the sample size, the width of the
resonances is not influenced by locally
varying $g$-factors. Therefore a inhomogeneous broadening of
the resonance line due to structural or chemical inhomogeneities
is avoided.  In this sense a Faraday-rotation fluctuation
measurement in the presence of an oscillating magnetic field
represent a continuous wave realization of an ESR spin-echo
experiment.

Such a measurement method could be usefull for samples
with weak structural homogeneity, such as
chemically synthesized $\rm CdSe$ quantum dots. In these
quantum dot arrays, the local variation of
the $g$-factors dominate inhomogeneous
broadening,\cite{Stern2005,Gupta2001} even for relatively
weak magnetic fields.
Unfortunately, the presence of an additional static field
component in $z-$direction leads to a shift of the spectrum
in Eq.~(\ref{result2}). Therefore inhomogeneous line broadening
due to hyperfine interaction persist also in the case of an
oscillating magnetic field.

While the physical limition of the proposed experiment
is that the sample structure must exhibit spin orbit
coupling and show the Faraday effect,
the generation of the oscillating magnetic field defines
the technical limitation. To resolve individual lines,
to measure the line width $T_2$, the separation of the lines
given by the field frequency $\gamma$ must exceed $T_2^{-1}$.
Further, to get a resonance at $\omega\neq 0$, the field
strength $\omega_0$ must be comparable to the field
frequency $\gamma$. With an experimentally challenging
magnetic field of the order of $10\,\rm mT$
at the frequency of the order $100\,\rm MHz$,
spin-coherence times down to some tens of ns
could be measured. With increasing spin-coherence times
the experimental requirements for the oscillating magnetic field
relax.  Therefore by using a Rb gas sample, with a spin-coherence time
exceeding $100\rm \mu s$,\cite{Crooker2004} such a multi-photon
absorption spectrum should be easily to realize.
It is worth to mention, that even if the spin decoherence time $T_2$
is significantly shorter than the field oscillation frequency, i.e.
if the multi-photon resonances are washed out, the dependence of the
spectrum on the  Bessel function still persists.\cite{Berns2006}

\begin{figure}[ht!]
\includegraphics[angle=-0,width=1\columnwidth]{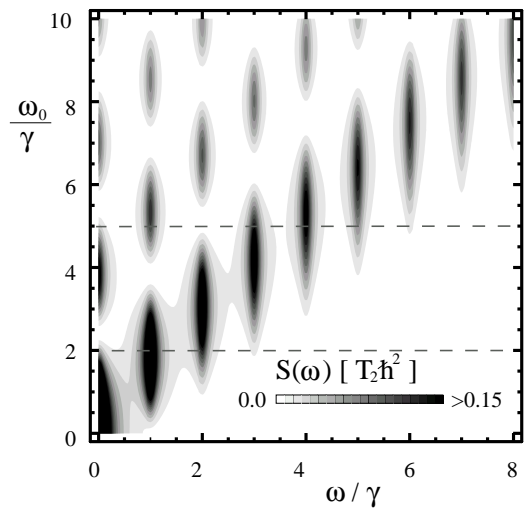}
\caption{\label{fig:densityplot}
Spin-spin correlation spectrum, as a function
of the ratio $\omega_0/\gamma$ and
frequency $\omega$ for $T_2^{-1}=0.1\gamma$.}
\end{figure}

In Fig.~\ref{fig:densityplot} we plot the spectrum of the
spin-spin correlation function $S(\omega)$
as function of $\omega/\gamma$ (noise frequency over field
frequency) and $\omega_0/\gamma$ (field strength over field frequency).
The two horizontal lines indicate the parameters
for the spectra shown in Fig.~\ref{fig:plotomega}.
If the time between the two spin measurements
gets significantly below the time needed by the
spins to close their trajectory, no spin correlation
can be measured. Therefore  Fig.~\ref{fig:densityplot}
shows no signal in the parameter range $\omega>\omega_0$.

Recently such a "Bessel staircase" as in Fig.~\ref{fig:densityplot}
was measured by Oliver {\em et al.}\cite{Oliver2005}
and  M. Sillanp\"a\"a {\em et al.},\cite{Sillanpaa2005} not in a
spin system but for the excitation probability in a strongly-driven
superconducting flux/charge qubit used as as Mach-Zehnder interferometer.
The main physical difference between Faraday-rotation fluctuation
spectroscopy and the Qubit case is, that in latter an average
quantity is measured, i.e. the probability for excitation, while
in former the correlation function is important. However, since
in lowest order perturbation theory, the
transition probability of the Qubit states can also be represented as
a correlation function,\cite{Berns2006} the close correspondence of
these two different experiments becomes immanent.

\section{\label{conclusions}Conclusions}

In conclusion, we present an elementary theoretical explanation of recent
experiments measuring spin noise in the presence of a static magnetic field
via the Faraday-induced fluctuations.
Furthermore, we predict the Faraday-rotation fluctuation spectrum if one
applies an oscillating magnetic field instead of a static one. Such an
experiment should offer an experimental easy possibility to observe
multi-photon absorption processes. Furthermore by such a measurement one
should be capable to measure spin-relaxation rates down to some
10th of ns, without inhomogeneous broadening due to structural or chemical
inhomogeneities, whereas the influence of hyperfine interaction will persist.

We thank D.~H\"agele, B. Kubala, M.~Oestreich,
and D. Urban for discussions. This work was
supported by the DFG under SFB 491 and GRK 726.

\end{document}